\documentclass[journal]{IEEEtran}
\ifCLASSINFOpdf
\else
\fi

\usepackage{setspace}
\usepackage{bbm}
\usepackage[cmex10]{amsmath}
\usepackage{amssymb}
\usepackage{cite}
\usepackage{graphicx}
\usepackage{array,color}
\usepackage{amsmath}
\usepackage{stfloats}
\usepackage{graphicx}
\usepackage{epstopdf}
\usepackage{subfigure}
\usepackage{tabularx}
\usepackage{epsfig,epsf,color,balance,cite}
\usepackage{algorithmic}
\usepackage{algorithm}

\ifodd 1
\usepackage{soul}
\usepackage{color}
\setstcolor{red}

\newcommand{\del}[1]{\st{#1}} 

\newcommand{\com}[1]{\textbf{\color{red} (COMMENT: #1)}} 
\newcommand{\response}[1]{\textbf{\color{green} (RESPONSE: #1)}} 
\else

\newcommand{\del}[1]{}

\newcommand{\com}[1]{}
\newcommand{\comg}[1]{}
\newcommand{\response}[1]{}
\fi

\begin{document}
\title{\huge{A Survey on Spatial Modulation in Emerging Wireless Systems: Research Progresses
		and Applications}}

\author{Miaowen Wen,~\IEEEmembership{Senior Member,~IEEE,} \normalsize Beixiong Zheng,~\IEEEmembership{Member,~IEEE,}
	Kyeong Jin Kim,~\IEEEmembership{Senior Member,~IEEE,}
	Marco Di Renzo,~\IEEEmembership{Senior Member,~IEEE,}
	Theodoros A. Tsiftsis,~\IEEEmembership{Senior Member,~IEEE,}\\
	Kwang-Cheng Chen,~\IEEEmembership{Fellow,~IEEE,}
	and Naofal Al-Dhahir,~\IEEEmembership{Fellow,~IEEE} \\
	\thanks{
		
		
		
		M.\ Wen and B.\ Zheng are with the School of Electronic and Information Engineering,
		South China University of Technology,
		Guangzhou 510641, China
		(email: eemwwen@scut.edu.cn; zheng.bx@mail.scut.edu.cn).
		
		K.\ J.\ Kim is with Mitsubishi Electric Research Laboratories, Cambridge,
		MA 02139 USA (e-mail: keyong.j.kim@hotmail.com).
		
		M.\ Di Renzo is with the Laboratoire des Signaux et Systemes,
		CNRS, CentraleSupélec, University Paris-Sud, Université Paris-Saclay,
		91192 Gif-sur-Yvette, France (e-mail: marco.direnzo@l2s.centralesupelec.fr).
		
		T.\ A.\ Tsiftsis is with the School of Intelligent Systems Science and Engineering,
		Jinan University, Zhuhai 519070, China (e-mail: theo\_tsiftsis@jnu.edu.cn).
		
		K.-C. Chen is with the Department of Electrical Engineering, University of
		South Florida, Tampa, FL 33620 USA (e-mail: kwangcheng@usf.edu).
		
		N. Al-Dhahir is with the Department of Electrical Engineering, The
		University of Texas at Dallas, Richardson, TX 75080 USA (e-mail:
		aldhahir@utdallas.edu).
	}
}

\maketitle
\begin{abstract}
Spatial modulation (SM) is an innovative and promising digital modulation technology that strikes an appealing trade-off between spectral efficiency and energy efficiency with a
simple design philosophy. SM enjoys plenty of benefits and
shows great potential to fulfill the requirements of future wireless communications. The key idea behind SM is to convey additional information typically through the ON/OFF states of transmit antennas and simultaneously save the implementation cost by reducing the number of radio frequency chains. As a result, the SM concept can have widespread effects on diverse applications and can be applied in other signal domains such as frequency/time/code/angle domain or even across multiple domains. This survey provides a comprehensive overview of the latest results and progresses in SM research. Specifically, the fundamental principles, variants of system design, and enhancements of SM are described in detail. Furthermore, the integration of the SM family with other promising techniques, applications to emerging communication systems, and extensions to new signal domains are also extensively studied.
\end{abstract}

\IEEEpeerreviewmaketitle

\section{Introduction}
Driven by the skyrocketing growth of mobile devices and the wide applications of the Internet of things (IoT), future wireless communication systems have triggered the explosive demand and urgent need for ultra-high capacity, ultra-low latency, and massive connectivity over the scarce wireless resources \cite{Lu2014An,Andrews2014What,wong2017key,Dahlman20145G}.
Specifically, an ever-increasing
volume of mobile data traffic is expected to appear in the coming years, which may overwhelm the limited spectrum resources significantly and increase the power consumption dramatically \cite{Andrews2014What,wong2017key,Dahlman20145G}. Due to the unprecedented surge of mobile data traffic,
researchers have been motivated to develop new transmission technologies for maximizing the achievable throughput and minimizing the deployment cost.
Among various technologies, spatial modulation (SM) has been envisioned as one prospective digital modulation technology to achieve high spectral efficiency and energy efficiency yet enjoy a simple design principle. Although early attempts, which can be dated back to the beginning of the 21-st century, have been made to explore the preliminary SM \cite{Mesleh2006Spatial,Mesleh2008Spatial}, they were sporadic and did not receive much attention in the early days.
After 2008, research work on SM has begun to grow explosively and the systematic introduction of the SM concept has been reported in some tutorial/overview literature \cite{Renzo2014Spatial,Renzo2011Spatial,Yang2015Design,Yang2016Single}, which attract a lot of attention from researchers.
By activating one transmit antenna to convey index information, SM reaps the spatial gain with a single radio frequency (RF) chain and
enjoys the following additional benefits \cite{Renzo2014Spatial,Mesleh2008Spatial}:
\begin{itemize}
	\item higher energy efficiency \cite{Stavridis2013Energy,Stavridis2012An};
	\item lower detection complexity with a smaller number of receive antennas and lower complexity of RF circuits;
	\item free of inter-channel interference;
	\item no need for inter-antenna synchronization;
	\item compatibility with massive multiple-input multiple-output (MIMO) \cite{Basnayaka2016Massive}.
\end{itemize}


In SM, extra information bits are typically conveyed through the index of one active transmit antenna, in addition to the information bits conveyed by the conventional constellation symbol.
Based on the antenna-switching mechanism, the active antenna index changes according to the spatial information bits.
In particular, only one RF chain is required at the SM transmitter to activate one out of multiple transmit antennas for a constellation symbol transmission, significantly saving the energy consumption in downlink communications, and dramatically reducing the hardware cost at the user terminal in uplink communications.
Moreover, due to freedom of the inter-channel interference, SM can be a better candidate technology than Vertical Bell Laboratories Layered Space�Time (VBLAST) for high-mobility wireless communication systems, where the channel correlation is weakened while the inter-channel interference effect is aggravated \cite{Cui2016Performance}.
Space shift keying (SSK) can be viewed as a simplified variant of SM \cite{Jeganathan2009Space,Renzo2012Space,Choi2016Sparse,Liang2016Coding-Aided,Renzo2010Improving,Renzo2011Bit11,Renzo2010Space}, which embeds the overall information into the index of one active antenna only without involving any constellation symbol.
Other variants of SM technology typically consist of generalized (G)SM \cite{Younis2010Generalised,Fu2010Generalised,Wang2012Generalised,Zheng2017Soft}, quadrature SM (QSM) \cite{Mesleh2015Quadrature}, differential (D)SM \cite{Bian2015Differential,Ishikawa2014Unified,Wen2015Low}, receive (R)SM \cite{Yang2011Transmitter,Stavridis2012Transmit,Zhang2013Generalised,Zhang2015Error}, and generalized (G)SSK \cite{Jeganathan2008Generalized}.
Generally speaking, SM refers to a new modulation family of communication systems that conveys additional information typically through the activation states of
transmit antennas.
 By choosing different activation patterns at the transmitter, various SM members provide a flexible design
 to meet different specific requirements and trade-offs among
 the spectral efficiency, energy efficiency, deployment cost, and system performance.
For information recovery, the receivers of SM have to execute two main tasks \cite{Basar2017Index}:
\begin{itemize}
	\item detecting the indices/states of active antennas;
	\item demodulating the constellation symbols embedded on the active antennas/states (if applicable).
\end{itemize}
However, it is not trivial to effectively detect both spatial and constellation information effectively
while maintaining low complexity for SM members under different channel conditions \cite{Zheng2017Soft,Zheng2017The}.
On the other hand, link-adaptive SM, which relies on the feedback from the receiver to alter its transmission pattern (e.g., modulation order, transmit power and antenna selection),
was extensively investigated to achieve better system performance and channel utilization \cite{Yang2011Adaptive,Yang2012Link,Yang2013Simplified,Yang2015Power}.
Please note that the detection performance of plain SM highly relies on the distinctness of the channel signatures/fingerprints associated with different transmit antennas. As a result, plain SM enjoys the technical challenge to operate in rich
scattering propagation and stationary \cite{Fu2016Performance} environments. With further leverage of advantage preprocessing techniques, such as orthogonal
pulse shaping \cite{Renzo2011Space} and trellis coded modulation (TCM) \cite{Mesleh2010Trellis,Basar2011New}, the problem of lack of channel distinctness can be
well addressed in SM systems.
Furthermore, as plain SM utilizes the space domain to convey index information via one active antenna, no transmit-diversity gain is provided by plain SM to combat channel fading effects.
To overcome the lack of transmit-diversity inherent in plain SM, transmit-diversity enhanced SM using space-time block coding (STBC) is also a promising direction to improve the error performance \cite{Basar2011Space-Time,Le2014Spatially,Li2014High,Jeon2015Multi-Strata,Renzo2013On}.

Due to its many promising advantages, SM serves as an attractive
energy-efficient modulation technology with flexibility of working together
with others in emerging communication systems.
Thanks to the sparsity inherent in SM signals, compressed-sensing (CS) theory is a powerful tool for low-complexity signal reconstruction even when the number of available measurements is much smaller than the signal dimension (i.e., the number of receive antennas is much less than that of transmit antennas in underdetermined systems), particularly in the case of large-scale MIMO \cite{Yu2012Compressed,Liu2014Denoising,Xiao2017Efficient,Wang2016NearML}.
On the other hand, due to the significantly reduced wavelength in the millimeter-wave (mmWave) frequency band, mmWave systems can be equipped with a large number of antennas at the transceiver in a highly compact manner for the implementation of large-scale MIMO \cite{Hur2013Millimeter}.
As a result, the SM family emerges as a promising low-cost and high-efficiency candidate for large-scale MIMO with a large number of transmit antennas while a much smaller number of RF chains are required for antenna activation.
More recently, noticing the great potentials of non-orthogonal multiple access (NOMA) in
supporting massive connectivity and low transmission latency, NOMA aided SM emerges as an attractive and novel technology for multi-user communications, which achieves high spectral efficiency and energy efficiency while maintaining low-complexity transceiver design \cite{zhu2017NOMA,Yang2019NOMA,Li2019Spatial}. As a result, NOMA aided SM technology
strikes an appealing trade-off among spectral efficiency, energy efficiency, deployment cost, and interference mitigation \cite{Zhong2018Spatial}. Simultaneous wireless information and power transfer (SWIPT) is another emerging important technology, which aims at delivering wireless information and energy concurrently. SM finds its special fit to SWIPT-enabled wireless systems since it has the potential to leverage the inactive antennas for energy harvesting without incurring any loss of the spectral efficiency.

Moreover, due to the broadcast nature of wireless communications, the security of SM transmission is an essential problem in practice \cite{Sinanovic2012Secrecy,Aghdam2015On,Guan2012On}. Interestingly, both physical layer security (PLS) and SM highly rely on the randomness and discrimination properties of the wireless interface, which can be exploited to achieve confidential information exchange among legitimate nodes while impairing the potential eavesdropper at the same time. Specifically, SM employs a fast antenna-switching mechanism to achieve the random selection of transmit antennas according to the spatial information bits and the legitimate channel state information (CSI), which can incur fast time-varying environments to confuse the eavesdropper and degrade its decoding performance.
In other words, SM shows a great potential not only in its increased energy efficiency, but also in the feasibility of secure transmission.
Furthermore, since visible light communications (VLC) enjoys high security and supports the fast switching in light-emitting diodes (LEDs),
SM technology can also be readily integrated into VLC, which provides the flexibility to control the luminance of the LED array and the communication throughput \cite{Fath2013Performance}.

Although originated in the space domain, the concept of SM is not exclusive to the space domain, but can be generalized and applied to other signal domains such as frequency/time/code/angle domain or even across multiple domains.
As a consequence, considerable interest and attention have been paid in recent years to fully develop various forms of the SM concept in
diverse wireless communication applications \cite{Basar2017Index,Ishikawa2018Years,Sugiura2017State}.
The use of the ON/OFF keying mechanism to embed index information has been mostly applied to entities in single domains such as spatial domain (e.g., antenna, RF mirror, and LED), frequency domain (e.g., subcarrier), time domain (e.g., time slot), code domain (e.g., spreading code and modulation type), and angle domain (e.g., angle of arrival (AoA) and polarization state).
To further enhance the system performance and enjoy a more flexible design, multi-dimension entities are also developed, which include more than one dimension for performing the ON/OFF keying mechanism.

The objective of this survey is to present a comprehensive overview of the latest results and progresses in SM research. In particular, Section~\ref{Variants of SM} reveals
the basic principles and variants of SM. In addition, Section~\ref{Enhancement SM} presents several performance enhanced techniques for SM systems, including link-adaptive SM, TCM aided SM and diversity enhanced SM.
The combinations of SM with other promising techniques and applications of SM to various emerging and practical communication systems are discussed in Sections~\ref{Combination SM} and \ref{Applications SM}. Finally, some important extensions of SM to other domains and entities are studied in Section~\ref{Extension SM}, and
concluding remarks with future directions are drawn in Section~\ref{Conclusions}.

\section{Basic Principle and Variants of SM}\label{Variants of SM}
In this section, we first introduce the basic principle of SM and then discuss some variants of SM, including (G)SM and (D)SM.
We consider the implementations of SM in a MIMO system with  $N_T$ transmit and $N_R$ receive antennas.
\subsection{Single-RF SM}

\begin{figure}[!t]
	\centering
	\includegraphics[width=3.5in]{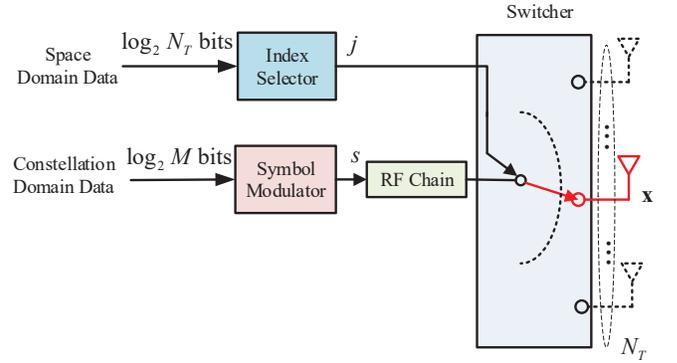}
	\caption{Transmitter diagram for the SM system.}
	\label{smtransmitter}
\end{figure}

At the beginning of this century, SM has emerged as a novel MIMO technology that works with a single RF chain and exploits the active antenna index to convey additional information based on the antenna-switching mechanism \cite{Mesleh2008Spatial,Renzo2014Spatial,Ntontin2013A}.
As a result, the information of SM is not only {\it explicitly} transmitted as one phase-shift keying (PSK)/quadrature amplitude modulation (QAM) symbol, but also {\it implicitly} transmitted by selecting the index of one active antenna for each channel use.
 Given the number of transmit antennas $N_T$ and the modulation order of the signal constellation $M$, the spectral efficiency of SM is
\begin{align}\label{SM_spectral}
S_{\text SM}=\log_2{N_T} + \log_2{M}~~\text{[bpcu]}
\end{align}
where $\text{[bpcu]}$ stands for bits per channel use.
Specifically, the first part of $\log_2{N_T}$ bits determines the index of
the active antenna $j$ and the second part of $\log_2{M}$ bits is used to modulate the
constellation symbol $s$. Consequently, as the constellation symbol $s$ is carried by the $j$-th antenna, the transmit vector of SM can be expressed as
\begin{align}\label{SM_signal}
{\mathbf{x}} = [ {\underbrace{0 \cdots 0}_{j-1} ~ {s} ~\underbrace{0 \cdots 0}_{N_T-j} } ]^T
\end{align}
where all elements of ${\bf x}$ except the $j$-th one are zeros.
As shown in Fig.~\ref{smtransmitter}, using the antenna-switching mechanism at the transmitter, the index of active antenna changes randomly for each channel use according to the incoming information bits.
Specially, when $M=1$, the SM system degenerates into the SSK system, in which the overall information bits are mapped into the index of active antenna in the space domain and the corresponding spectral efficiency is
\begin{align}\label{SSK_spectral}
S_{\text SSK}=\log_2{N_T} ~~\text{[bpcu]}.
\end{align}

At the receiver side, both the index of active antenna and the corresponding constellation symbols must be detected for information recovery. Various detection algorithms have been developed for SM systems, which can be divided into two main categories: joint detection and decoupled/separate detection.
For example,
the maximum-likelihood (ML) detector is the most typical representative of joint detection, which searches jointly over all possible transmit antennas and constellation symbols
to achieve optimal detection performance \cite{Jeganathan2008Spatial}.
However, the ML detector has relatively high complexity which linearly scales with the number of transmit antennas and the size of the constellation.
On the other hand, various suboptimal lower-complexity detectors were also developed, which typically involve a two-step decoupled process, i.e.,
first determine the index of the active transmit antenna and then demodulate the constellation symbol carried on the active transmit antenna \cite{Naidoo2011Spatial,Sugiura2011Reduced-Complexity,Tang2013A,Zheng2012Signal,Li2015Low-Complexity,Maleki2016On}. As a result, the size of the
search space is reduced from $\mathcal{O}(N_T*M)$ to $\mathcal{O}(N_T+M)$.
Although
the two-step decoupled/separate detectors have smaller search complexity,
they usually suffer from performance degradation compared
with the optimal detector.
Therefore, it is an essential and challenging problem to design low-complexity detection algorithms to achieve optimal or near-optimal performance for the SM receiver.
It is noted that by exploiting the sparsity of the SM signal, the implementations of low-complexity detection algorithms based on CS theory are developed to achieve near-optimal performance \cite{Yu2012Compressed,Garcia-Rodriguez2015Low-Complexity}.

On the other hand, the practical shaping filter is employed to evaluate the impact of antenna
switching for band-limited SM and a multiple-RF antenna switching mechanism is developed to overcome the shortage of single-RF SM scheme in \cite{Ishibashi2014Effects}.
Moreover, the authors of \cite{Arisaka2019Energy} evaluate the impact of pulse shaping on the bandwidth-versus-energy-efficiency tradeoff for band-limited SM and emphasize the advantage of single-RF SM in large-scale antenna array implementation.
Moreover, the design of SM detection algorithms taking into account of channel correlation \cite{Renzo2010Performance,Chang2013Detection}, imperfect channel estimation \cite{Rajashekar2014Reduced-Complexity,Sugiura2012Effects,Mesleh2012On,Wu214Channel},
performance analysis \cite{Renzo2012Bit,Renzo2011Bit22,Yang2016Transmit,Younis2013Performance},
cooperative protocol \cite{Narayanan2016Distributed,Zhang2016On,Yang2016Spatial},
 and coded systems \cite{Tang2013A,Xiang2017A,Li2015Low-complexity1,Serafimovski2010Fractional} are also extensively investigated in the literature.

It is also worth pointing out that the concept of SM provides a multiple-antenna full-duplex node additional freedom to decide which antennas for signal transmission and which antennas for reception simultaneously, increasing the spectral efficiency while remaining the RF chains advantages. The authors of \cite{Jiao2014Spatial} for the first time proposed this idea, which was later extended to full-duplex relaying, such as in \cite{Raviteja2016Spatial}.
Note that in the above introduction the number of transmit
antennas $N_T$ has been assumed to be a power of $2$. When
this assumption cannot be satisfied, effective schemes such as
fractional bit encoding \cite{Serafimovski2010Fractional}, bit padding \cite{Yang2011Bit}, and constellation
order varying \cite{Yang2012Information} can be applied to improve the spectral
efficiency of the SM system. The investment for the benefit is
the increased susceptibility to error propagation.


\subsection{Generalized (G)SM}
Although the single-RF SM enjoys high energy efficiency with one active antenna, one major disadvantage is that spectral efficiency suffers from a slow logarithmic growth with an increasing number of transmit antennas.
Hence, the constraint of a single active antenna
is relaxed in (G)SM to provide more spatial information by allowing more than one antennas to be simultaneously
activated to transmit the same PSK/QAM symbol \cite{Younis2010Generalised,Fu2010Generalised}.
Since the same PSK/QAM symbol is transmitted from all active antennas,
this (G)SM scheme requires only one RF chain and
is also free of inter-channel interference and there is no need for inter-antenna synchronization.
For each channel use, $K$ out of $N_T$ ($K \le N_T$) transmit
antennas are selected to carry the same constellation symbol while the remaining $N_T-K$ antennas are inactivated, resulting in the following spectral efficiency\footnote{${{N_T}\choose{K}}$ stands for the binomial coefficient and $\left\lfloor\cdot\right\rfloor$ is the floor function.}
\begin{align}\label{GSM1_spectral}
S_{\text GSM1}=\left\lfloor {\log_2}{{N_T}\choose{K}} \right\rfloor + \log_2{M}~~\text{[bpcu]}.
\end{align}
Although the spatial information significantly increases in (\ref{GSM1_spectral}) compared to (\ref{SM_spectral}), the constellation information is limited by one PSK/QAM symbol.
To further improve the spectral efficiency, the concept of (G)SM can be further extended
by allowing different active antennas to carry different PSK/QAM symbols \cite{Wang2012Generalised,Xiao2017Time-Domain}, resulting in the spectral efficiency
\begin{align}\label{GSM2_spectral}
S_{\text GSM2}=\left\lfloor {\log_2}{{N_T}\choose{K}} \right\rfloor + K\log_2{M}~~\text{[bpcu]}.
\end{align}
In particular, (G)SM enables a flexible design to achieve a compromise between the
spectral efficiency, the deployment cost (the number of RF chains), and the
error performance by choosing the required number of active antennas \cite{Patcharamaneepakorn2016Spectral}.
It is worth pointing out that the single-RF SM and full activation MIMO become two special cases of (G)SM with $K=1$ and $K=N_T$, respectively.

\begin{figure}[!t]
	\centering
	\includegraphics[width=3.5in]{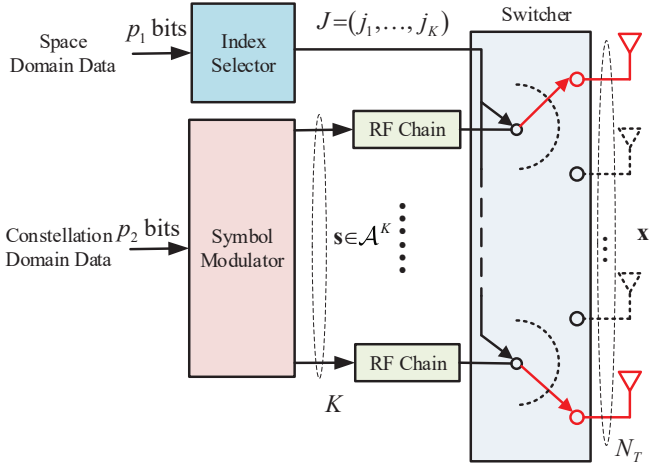}
	\caption{Transmitter diagram for the (G)SM system.}
	\label{gsmtransmitter2}
\end{figure}
The transmitter diagram of the (G)SM system is depicted in Fig.~\ref{gsmtransmitter2}, where the incoming information bits are separated into two parts for different purposes.
Specifically, unlike the classical MIMO scheme that maps all incoming bits to constellation points for all transmit antennas,
a block of $p=p_1+p_2$ incoming bits is divided into the space domain and constellation domain, respectively.
The first part with $p_1=\left\lfloor {\log_2}{{N_T}\choose{K}} \right\rfloor$ bits is applied to the index selector for activating $K$ antennas (i.e., determining the indices of those active antennas)
\begin{align}\label{indices}
J = \left( j_1, j_2, \ldots ,j_{K} \right)
\end{align}
where ${j_k } \in \mathcal{T}\triangleq \left\{ {1, \ldots ,{N_T}} \right\}$ is the index of the $k$-th active antenna with $j_1< j_2 \cdots< j_{K}$ and
$J$ stands for the active antenna combination (AAC), whose bit mapping procedure can be implemented using the look-up table when the number of AACs is small or combination strategy when the number of AACs is large \cite{Basar2013Orthogonal,Knuth2005The,Narasimhan2016On}.

It is worth noting that among ${{N_T}\choose{K}}$ possible combinations, only $N_L = 2^{p_1}$ AACs are permitted while the other ${{N_T}\choose{K}}-N_L$ AACs are illegal due to the integer constraint of mapping bits.
Without loss of generality, the set of $N_L$ legal AACs is denoted as ${\cal{J}}=\left\{{J}_\jmath\right\}_{\jmath=1}^{N_L}$.
The second part with $p_2=K\log_2 M$ bits is then applied to the symbol modulator for generating $K$ constellation symbols transmitted by the active antennas, which generates the transmitting block as
\begin{align}\label{GSM transmission block}
x_n=\left\{ \begin{gathered}
s_k,~n = j_k, \hfill \\
0 ,~~ n  \notin J, \hfill
\end{gathered}  \right.\quad n=1,\ldots,N_T
\end{align}
where $s_k$ is drawn from an order-$M$ constellation ${\cal{{A}}}=\left\{a_\imath\right\}_{\imath=1}^{M}$ and transmitted by the $k$-th active antenna.
Denote ${\bf x} \triangleq \left[x_1~ x_2 \ldots x_{N_T}\right]^{T} $, where
$x_n \in \{0\} \cup\cal{{A}}$ and $\lVert{\bf x}\rVert_0=K$.

\begin{figure}[!t]
	\centering
	\includegraphics[width=3.5in]{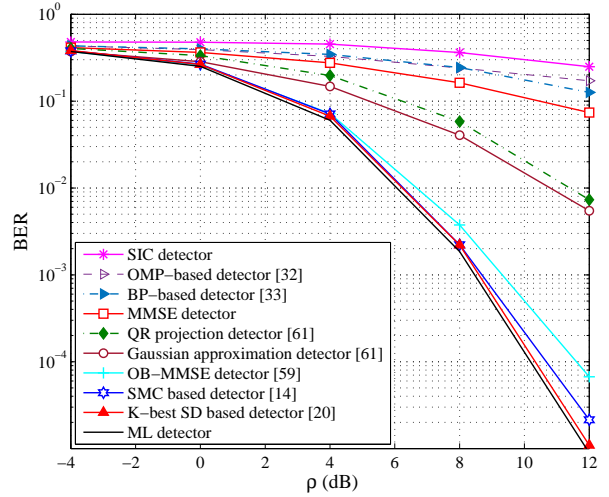}
	\caption{BER performance comparison of different detectors in the uncoded (G)SM system
		with $N_T=N_R=8$, $K=4$, and QPSK modulation. Simulation results extracted
		from \cite{Zheng2017Soft} and \cite{Zheng2017The}}.
	\label{uncoded GSM8_8}
\end{figure}
\begin{figure}[!t]
	\centering
	\includegraphics[width=3.5in]{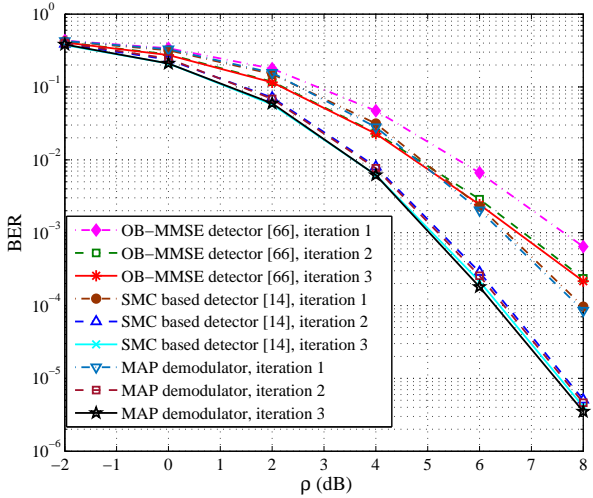}
	\caption{BER performance comparison in the coded (G)SM system with $N_T=N_R=8$, $K=4$, and QPSK modulation, where the iterative (G)SM receiver and rate-1/2 constraint length-3 convolutional code are employed. Simulation results extracted
		from \cite{Zheng2017Soft}.}
	\label{Turbo8}
\end{figure}
After transmission through the MIMO channel, the signal model at the receiver side is expressed as
\begin{align}
{\bf{y}} &=\sqrt{\frac{\rho}{K}} {\bf{Hx}} + {\bf{w }}
 =\sqrt{\frac{\rho}{K}} {{\bf{H}}_{J}}{\bf{s}} + {\bf{w}}\label{rec_vector2}
\end{align}
where $\rho$ strands for the transmit power, ${\bf{w}}\sim \mathcal{N}_c({\bf 0}, {\bf I}_{N_R})$ denotes an additive white Gaussian noise (AWGN) vector with a zero mean and normalized variance, ${\bf{s}}\triangleq \left[s_{1}~ s_{2}\ldots s_{K}\right]^{T} \in {\cal{{A}}}^K$  is a $K \times 1$ constellation symbol vector carried by the active antennas, and ${\bf{H}}_{J}\triangleq\left[{\bf{h}}_{j_1}~ {\bf{h}}_{j_2} \ldots {\bf{h}}_{j_K}\right]$ consists of the $K$ columns of ${\bf{H}}$ corresponding to $K$ active transmit antennas.
The receiver of (G)SM has to
detect the indices of the active antennas
and demodulate the corresponding constellation symbols simultaneously for information recovery.

By performing a joint detection over the active antenna indices and constellation symbols, the ML detector for (G)SM can be expressed as
\begin{align}\label{MLdetector}
\left\langle {{\hat J},{\hat{\bf s}}} \right\rangle =\mathop {\arg \min }\limits_{{J\in {\cal{J}}},~{{\bf{s}}}\in {\cal{A}}^K}
\left\|{\bf{y}} -\sqrt{\frac{\rho}{K}} {{\bf{H}}_{J}}{\bf{s}}\right\|^2.
\end{align}
We can observe from (\ref{MLdetector}) that the size of the search space is $\mathcal{O}(N_L\cdot M^K)$, resulting in a considerably high
search complexity, especially for large-scale MIMO or high order constellations.
Therefore, many researchers have
 focused on the design of low-complexity detection algorithms for the (G)SM system, e.g., the CS based detections \cite{Yu2012Compressed,Liu2014Denoising,Xiao2017Efficient,Wang2016NearML}, the ordered block minimum mean-squared-error (OB-MMSE) based detections \cite{Xiao2014Low,Chen2015GSMdetect}, the message passing based detection \cite{Narasimhan2016On}, the Gaussian approximation based detections \cite{Lin2015GSM,He2015Bayesian}, and
 sphere decoding (SD) based detections \cite{Cal-Braz2014Sphere,Younis2013Tcom,Younis2011ICC}.
 However, the above-mentioned detection algorithms either suffer from suboptimal performance with considerable loss or
 encounter a fluctuating complexity that depends on the channel conditions, making them unsuitable for practical implementations.
Moreover, most existing detectors output hard-decisions, which suffer from some performance loss when channel codes are deployed in the coded (G)SM system.
In \cite{Xiao2016SoftGSM}, a soft OB-MMSE based detector is proposed to output soft-decisions for the coded (G)SM system,
which sorts the possible AAC based on the specific metric and then proceeds to estimate the constellation symbols associated with the MMSE solution for each AAC.
To achieve near-optimal performance but with low complexity for coded (G)SM systems, soft detection algorithms based on
deterministic sequential Monte Carlo (DSMC) and $K$-best SD
are proposed in \cite{Zheng2017Soft,Zheng2017The}, respectively.
Fig. \ref{uncoded GSM8_8} shows the performance results of some representative detectors in terms of bit error rate (BER) for the uncoded (G)SM system.
It can be observed from Fig. \ref{uncoded GSM8_8} that by showing the ML detector as the benchmark,
the detectors proposed by \cite{Zheng2017Soft} and \cite{Zheng2017The} can achieve near-optimal performance while other detectors still suffer
from some performance loss.
To examine the performance of soft decoders, Fig. \ref{Turbo8} shows the comparison results in terms of BER for the coded (G)SM system, where the turbo receiver and rate-1/2 constraint length-3 convolutional code are employed.
It can be observed that the DSMC based detector in \cite{Zheng2017Soft} achieves nearly the same performance as the MAP detector (acts as the performance benchmark for the turbo receiver) under different iteration numbers, which outperforms the soft OB-MMSE detector in  \cite{Xiao2016SoftGSM} significantly.

\subsection{Differential (D)SM}

Since the SM transmitter embeds the spatial information in the active antenna implicitly,
the SM receiver has to detect such spatial information by distinguishing different channel fading states associated with different transmit antennas, which requires the CSI for coherent detection. However, the CSI requirement at the receiver increases the deployment cost due to the pilot overhead and channel estimation complexity.
Alternatively, differential encoding of the SM symbols, which dispenses with any CSI at the transceiver while inheriting the advantages of SM, emerges as an attractive solution with low deployment cost. A differentially encoded space-time shift keying (STSK) modulation scheme is presented in \cite{Sugiura2010Coherent} as the primitive idea of (D)SM, which employs the Cayley unitary transform and conveys information via the activation state of the space-time dispersion matrix.
To overcome limitation on the real-valued constellation of \cite{Sugiura2010Coherent},
the authors of \cite{Sugiura2011Reduced-Complexity} further develop a
differentially STSK scheme with QAM to achieve higher bandwidth-efficiency.

\begin{figure}[!t]
	\centering
	\includegraphics[width=3.5in]{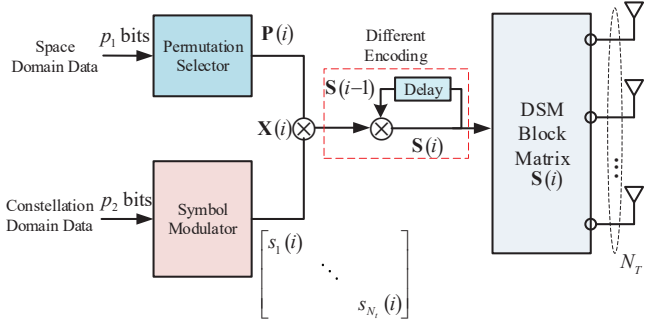}
	\caption{Transmitter diagram for the (D)SM system.}
	\label{dsmtransmitter}
\end{figure}

\begin{figure}[!t]
	\centering
	\includegraphics[width=3.5in]{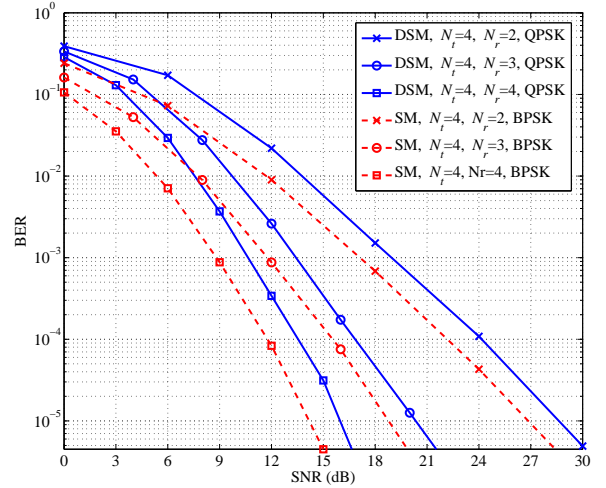}
	\caption{BER performance of (D)SM versus coherent detection SM at $3$~bps/Hz transmission rate with $N_T=4$.}
	\label{dsmfig03}
\end{figure}

\begin{figure}[!t]
	\centering
	\includegraphics[width=3.5in]{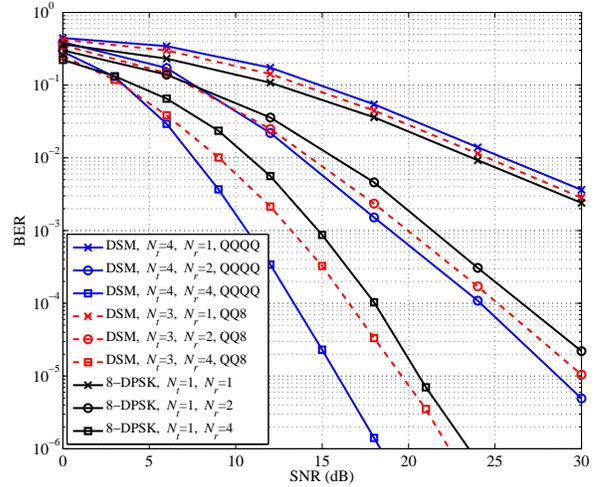}
	\caption{BER performance of (D)SM with $N_T=3$ and $N_T=4$ at $3$~bps/Hz transmission rate, where ``Q" and ``8" in the legend denote QPSK and 8PSK,
		respectively.}
	\label{dsmfig04}
\end{figure}

On the other hand, a permutation-based (D)SM scheme is developed in \cite{Bian2015Differential}, which conveys information via the
antenna activation order and is applicable to an arbitrary number of transmit antennas.
Specifically, as the original SM of $N_T$ antennas has $N_T$ activation states, the (D)SM
will transmit an $N_T \times N_T$ space-time block, which is one
permutation of block activation states. As a result, we have $N!$ different
permutations for the space-time block and the transmitter of (D)SM selects one out of $N_T !$ permutations according to the previous space-time block and the incoming spatial information bits, which is illustrated in Fig. \ref{dsmtransmitter}.
For each (D)SM block, the first part with $p_1=\left\lfloor {\log_2}({{N_T}!)} \right\rfloor$ bits is applied to the permutation selector to determine a permutation ${\bf P}(i)$, while the second part with $p_2=N_T\log_2 M$ bits is then applied to the symbol modulator for generating $N_T$ constellation symbols in the form of the diagonal matrix ${\text {diag}}\{s_1(i) \cdots s_{N_T}(i)\}$, resulting in the information block
 \begin{align}\label{DSM1}
 {\bf X}(i)={\text {diag}}\{s_1(i) \cdots s_{N_T}(i)\}{\bf P}(i)
 \end{align}
 where ${\text {diag}}(\cdot)$ denotes the diagonal operation. Finally, the (D)SM block matrix is calculated by
 \begin{align}\label{DSM2}
 {\bf S}(i)={\bf S}(i-1){\bf X}(i)
 \end{align}
 where the initial block matrix
 of (\ref{DSM2}) can be assumed to be the identity matrix, i.e., ${\bf S}(0)={\bf I}_{N_T}$.
As it can be seen, the $N_T \times N_T$ space-time block is a full-rank matrix and has only one non-zero entry in each row or column.
By leveraging the time-coherent property for non-coherent detection,
the (D)SM can successfully avoid the requirement of pilot insertion at the transmitter and channel estimation at the receiver to achieve low deployment cost. Fig.~\ref{dsmfig03} shows the BER comparison results between  non-coherent detection (D)SM and coherent detection SM with $N_T=4$, targeting at the spectral efficiency of $3$ bps/Hz. It can be observed that compared with coherent detection SM,
the performance loss in terms of BER
for the non-coherent detection (D)SM is less than $3$ dB.
 Fig.~\ref{dsmfig04} shows the BER performance results of (D)SM under different parameter settings,
 targeting at the spectral efficiency of $3$ bps/Hz.
 By showing the single-antenna differential PSK (DPSK) as a benchmark,
it can be observed that the performance gain of (D)SM increases as the number of receive antennas grows.


A reduced complexity detector is investigated for (D)SM in \cite{Wen2015Low} to achieve near-optimal performance, which
is derived from the ML principle and adopts a two-step operation.
The introduction of amplitude phase shift keying (APSK) modulation to
(D)SM is investigated in \cite{Martin2015Differential} and \cite{Liu2017High} to enhance the transmission efficiency and performance.
In \cite{Wen2014Performance}, the performance of (D)SM with two transmit antennas over Rayleigh fading channels is analyzed in terms of the average bit error probability, which shows a less than 3 dB performance loss compared with the coherent SM. To fill the performance gap between the DSM and the coherent SM, the state-of-the-art in rectangular DSM concept is extensively investigated in \cite{Ishikawa2017Rectangular,Wu2018Space,Wu2019Space,xiao2019rectangular,Ishikawa2018Differential}.

On the other hand, DSM also has various applications.
For example, the concept of (D)SM is then implemented in dual-hop networks \cite{Zhang2016Dual}, which reaps the transmit-diversity and reduces the receive complexity by using precoding.
To achieve the improved BER performance of (D)SM, the concept of Gray coding is applied, where the Trotter-Johnson ranking and unranking theory is adopted to perform index permutations \cite{Li2016Differential}.
To capture the full transmit-diversity of (D)SM, schemes that use cyclic signal constellation and algebraic field extensions are developed in \cite{Zhang2015Differential} and \cite{Rajashekar2017Full}, respectively.
Inspired by the (G)SM concept, the authors of \cite{Ishikawa2014Single} generalize the differential STSK scheme
by allowing a subset of space-time dispersion matrices to carry more than one real-valued constellations.
Furthermore, the differential transmission scheme for (G)SM with multiple active transmit antennas is described in \cite{Helmy2017Differential}, which strikes a more flexible trade-off between the diversity gain and transmission rate.

\subsection{Receive (R)SM}
As a reciprocal scheme of SM technology, RSM, a.k.a. precoding aided SM (PSM), has also received a lot of research interest due to its simplified receive structure. Specifically, by applying the concept of SM at the receiver-side, RSM utilizes the indices of receive antennas to convey spatial information in addition to conventional constellation information of PSK/QAM symbol. With the aid of transmitter precoding, RSM benefits from both high beamforming gain and low complexity design at the receiver side, which is highly desired for the downlink MIMO transmission.

With the perfect CSI available at the transmitter, the initial RSM is explored in \cite{Yang2011Transmitter}, where both zero-forcing (ZF) and minimum mean-squared error (MMSE) pre-coding schemes are considered.
Two precoding schemes for RSM under imperfect CSI at the transmitter-side are then developed in \cite{Stavridis2012Transmit}.
Moreover, the concept of RSM is also generalized by activating more than one antennas at the receiver-side \cite{Zhang2013Generalised,Zhang2015Error}, in which error performance and the low-complexity detectors are investigated.
In \cite{Zhang2016Performance}, a non-linear RSM scheme is developed with the leverage of vector perturbation, which conveys implicit information via the activation pattern at the receiver-side.
The authors of \cite{Liu2018Transmitter} develop a new RSM scheme to achieve both transmit and receive diversity, where several associated detection algorithms are introduced to meet different requirements in terms of complexity and reliability.
To reduce the significant channel estimation overheads required at the transmitter, a two-stage RSM based on partial CSI is proposed in \cite{Maleki2019Receive} for correlated channels.
Moreover, to overcome the performance degradation due to the channel correlation, an orthogonality structure is designed for generalized RSM \cite{Cao2019Orthogonality}.
Based on the ML criterion, both coherent and incoherent detection schemes are studied in \cite{Mokh2018Theoretical}, which are then simplified to the single-tap detection problem.

More recently, the performance of RSM in various application scenarios are also studied and discussed.
An upper bound on the error performance of the MMSE-based RSM system is derived in \cite{Li2018Performance}.
In \cite{Yang2018Unified,Yang2019Enhanced,Li2018Power}, the power allocation problem for RSM is formulated, where new approximate solutions are proposed to achieve a higher performance gain.
The error performance, diversity order, and coding gain of RSM under the shadowing MIMO channel are analyzed in \cite{Stavridis2017Performance} and \cite{Stavridis2016On}.
In \cite{Stavridis2016Performance}, the integration of multi-stream RSM under a broadcast channel is studied for the multi-user scenario, where the error performance, diversity order, and coding gain are derived.
The authors of \cite{LiGeneralized2017} develop an amalgamated scheme of QSM and RSM and derive a closed-form upper bound on its error performance.
Moreover, the designs of RSM for secure communication are also considered in \cite{Chen2016Secure,Wu2015Secret,Wu2016Transmitter}.
In \cite{Cheng2017On}, the integration of RSM into the simultaneous wireless information and power transfer system is studied, where the rate-energy trade-off problem is revealed with performance optimization.

\section{Performance Enhancement for SM}\label{Enhancement SM}
As Section~\ref{Variants of SM}, the system performance of SM and its variants highly depends on the distinctness of the channel signatures/fingerprints associated with different transmit antennas. To avoid significant performance degradation, those SM members require rich scattering
in the propagation environment, which, however, may not be realistic in practical systems.
Fortunately, with preprocessing techniques, such as link-adaptive, precoding/TCM, and space-time-coded transmission, the error performance of the SM family can be improved significantly in adverse environments.
\subsection{Link-Adaptive SM}
Under the slow fading channel model, link-adaptive SM can be applied  to achieve better BER performance and higher channel utilization by adapting the parameter settings (e.g., modulation order, transmit power and number of transmit antennas) or precoding matrices according to the channel condition.
Specifically, after channel estimation, the receiver determines the optimal transmission parameters in terms of modulation order, power allocation profile and antenna selection mode, or the optimal precoders, which is then sent back to the transmitter using a limited feedback link \cite{Yang2011Adaptive,Yang2012Link,Yang2013Simplified,Yang2015Power}.

In \cite{Bouida2015Adaptive}, adaptive SM mechanisms, which adapt the modulation type and/or transmit power
through the limited reliable feedback channel, are introduced to cognitive radio systems to enhance both the spectral and energy efficiency of the secondary system.
An adaptive SM scheme based on the Huffman coding technique is proposed in \cite{Wang2017Huffman}, which activates transmit antennas with distinct probabilities via Huffman mapping with a variable-length code. In addition, \cite{Maleki2014Adaptive} exploits spatial correlation as the partial CSI at the transmitter to realize the adaptive SM principle, which is shown to achieve performance improvement.
An adaptive unified linear precoding scheme based on the maximum criterion of the minimum Euclidean distance for (G)SM systems is proposed in \cite{Cheng2018A}, which maintains the RF chains and is shown to improve the system error performance relative to the (G)SM systems without precoding.
The adaptive SM principle is also introduced in mmWave communication systems, where the shortest Euclidean distance of the
SM symbols is maximized in an iterative manner to improve error performance \cite{Luo2017Adaptive}.
\subsection{Precoding/TCM Aided SM}
It is worth pointing out again that the detection performance of the spatial information highly depends on the distinctness of channel impulse responses associated with different transmit antennas.
As a result, the detection performance suffers from spatially-correlated fading channels as the channel impulse responses become very similar from different transmit antennas.

The precoding and TCM are two effective techniques to overcome the negative impact of spatial correlation on SM.  In \cite{Koca2018Precoding}, a diagonal precoding matrix is applied to the SM transmit vector before transmission, which is designed to minimize the asymptotic average BER while preserving the average power budget and without any explicit knowledge of the channel coefficients at the transmitter. On the other hand, in \cite{Mesleh2010Trellis} and \cite{Basar2011New} the encoding scheme based on
TCM is investigated for SM systems. Specifically, the
spatial information bits are convolutionally encoded and interleaved according to the TCM principle in \cite{Mesleh2010Trellis}, which differentiates the transmit antennas into multiple subsets to maximize the spatial distance between each transmit antenna within the same subset.
However, although the scheme proposed in \cite{Mesleh2010Trellis} benefits from some performance improvements in correlated fading channels, such improvements vanish in uncorrelated fading channels, compared with the conventional SM without TCM. To break the bottleneck of \cite{Mesleh2010Trellis}, another TCM based SM scheme is proposed in \cite{Basar2011New}, which reaps the advantages of trellis coding and achieves performance improvement over the conventional SM in both uncorrelated and correlated fading channels.
\subsection{Transmit-Diversity Enhanced SM}
It has been reported in \cite{Renzo2014Spatial} that no transmit-diversity gain can be achieved by SM transmission even though the transmitter is equipped with multiple antennas. Hence, many researchers tried to develop transmit-diversity techniques with encoding mechanisms to overcome the lack of transmit-diversity inherent in SM.

In \cite{Basar2011Space-Time}, the Alamouti code is exploited as the STBC to integrate with SM, which reaps the second-order transmit-diversity and enjoys the system simplicity by activating a pair of transmit antennas. Moreover, owing to the orthogonality of the Alamouti STBC on a
spatial-constellation diagram, an optimal detector is further developed for the STBC-SM system with linear complexity.
By resorting to the non-vanishing determinant property, the orthogonal STBC is embedded into the spatial constellation to achieve the transmit-diversity order of two \cite{Le2014Spatially}. However, although the orthogonal STBC based SM scheme in \cite{Le2014Spatially}
achieves higher spectral efficiency without the rotation phase optimization involved in \cite{Basar2011Space-Time}, it is limited by its antenna configuration, which requires that the number of transmit antennas should be even and not less than four.
To overcome the spectral efficiency degradation in the STBC-SM system of \cite{Basar2011Space-Time}, an enhanced STBC-SM making use of the cyclic structure and signal constellation rotation is developed in \cite{Li2014High}, which improves the spectral efficiency without scarifying the transmit-diversity. By optimizing the power and phase allocation, a SM scheme based on modified multi-strata STBC is also proposed to further improve the spectral efficiency while maintaining the same transmit-diversity order \cite{Jeon2015Multi-Strata}.
In \cite{Helmy2016Enhanced}, temporal permutations are introduced as a new dimension to improve the performance of spatially modulated STBCs and ensure a full transmit diversity order.

To enjoy the high transmit-diversity order while avoiding channel estimation, the STBC is also integrated into the (D)SM, which reaps the diversity order without the requirement of CSI at the receiver \cite{Xiao2017Space-Time}.
Inspired by the QSM proposed in \cite{Mesleh2015Quadrature},
the in-phase and quadrature parts of the constellation symbols are
linearly multiplexed by  a pair sets of dispersion matrices in \cite{Wang2017Diversity-Achieving} to obtain the diversity-achieving QSM transmission mechanism, which achieves the second-order transmit-diversity and improves the spectral efficiency by using
additional spatial information embedded in a dispersion matrix associated with the quadrature part.

\section{(G)SM Integration with Other Promising Technologies}\label{Combination SM}

\subsection{Compressed-Sensing (CS) Theory for SM}

For (G)SM systems, conventional detection methods such as ML and
linear detections can be applied. However, the ML detector has prohibitively high complexity
that exponentially grows with the number of active antennas whereas linear detectors, e.g., MMSE and ZF suffer from significant performance loss. Recently, by exploiting the sparse features in (G)SM symbols, several researchers focused on the design of low-complexity CS-based detection schemes to achieve satisfactory performance.
Therefore, CS theory becomes a powerful tool that makes the signal reconstruction possible even when the number of available measurements is much smaller than the signal dimension (i.e., the number of receive antennas is much less than that of transmit antennas in underdetermined systems),
particularly in the case of large-scale MIMO \cite{Chen2019Off}.
Since a very large number of transmit antennas are used in large-scale MIMO while
only a few of them are active during each transmission, the (G)SM
symbols can be considered sparse since most elements of the transmitted block are zeros.

By formulating the antenna detection of SM as an $\ell_1$-norm convex optimization problem, \cite{Yu2012Compressed} applies the normalized CS to detect the active antenna indices, which saves the computational complexity significantly in the large-scale MIMO implementation.
Against the limitation that basis pursuit (BP) algorithm is not suitable for detecting sparse signals corrupted by noise, \cite{Liu2014Denoising} makes some adaptations to counter the effect of noise and formulates the detection of (G)SM signals as a convex quadratic programming problem. A sparse Bayesian learning based detection algorithm is then proposed in \cite{He2015Bayesian} to reconstruct the (G)SM symbol effectively with low complexity, which shows outstanding performance over other CS-based schemes in the highly underdetermined scenarios (i.e., the receiver only has a small number of antennas).
 By formulating the (G)SM detection as an $\ell_0$-norm optimization problem under the constraint
of fixed sparsity, Bayesian CS is applied in \cite{Wang2016NearML} to solve the detection problem, which shows a better performance than the BP based detector.
Several efficient CS-based detectors under the orthogonal matching pursuit framework are proposed in \cite{Xiao2017Efficient} for large-scale (G)SM systems, which achieve satisfactory performance with considerably reduced complexity.

The application of CS theory to the multiuser detection problems of SM systems also received great attention from many researchers.
In \cite{Garcia-Rodriguez2015Low-Complexity}, the detection problem for SM systems over the multiple-access channel is investigated, in which a low-cost detector based on CS theory is proposed by exploiting the structured sparsity of the problem.
Exploiting the distributed and group sparsity in large-scale SM uplink systems, \cite{Gao2016Compressive-Sensing-Based} develops a structured CS-based multiuser detection algorithm with satisfactory performance and low deployment cost.
The joint detection problem of user activity and (G)SM signal is investigated over the terrestrial return channel \cite{Wang2018Generalized}, in which the block-sparse CS-based method is developed
for solving the joint detection problem by exploiting the structured sparsity inherent in multi-user (G)SM systems.
In \cite{Xiao2018Compressed-Sensing}, after partitioning transmit antennas into multiple groups to perform independent SM,
corresponding detectors based on threshold-aided CS and message passing are proposed to achieve near-optimal performance with low complexity.
In \cite{Xiao2018Bayesian}, an effective Bayesian CS detector is proposed with low deployment cost for STBC based QSM systems.



\subsection{Non-orthogonal Multiple Access (NOMA) Aided SM}
NOMA, which exhibits great potential in supporting both massive connectivity and low transmission latency, has been envisioned as one of the most promising
technologies to accommodate more active users
in the future network \cite{Dai2015mag}. In NOMA networks, multiple users share the same channel resource and are multiplexed in the power domain via superposition coding at the transmitter \cite{Ding2017mag}. To mitigate inter-user interference, multi-user detection (MUD) techniques such as successive interference cancellation (SIC) are typically employed at the receiver.
On the other hand, most research works of SM focus on the point-to-point transceiver design and current multi-user SM systems usually suffer from inter-user interference \cite{serafimovski2012multiple}.
Recalling that NOMA can effectively mitigate inter-user interference, NOMA aided SM emerges as a competitive technology in multi-user communications
to achieve high spectral efficiency and energy efficiency yet maintain low deployment cost and reduced inter-user interference.
In particular, NOMA-SM strikes an appealing trade-off between spectral efficiency, energy efficiency, and interference mitigation.
Towards this direction, researchers are motivated to promote the deployment of NOMA aided SM and aim to explore both the potential benefits of SM with low-complexity design and NOMA with high bandwidth efficiency.

In \cite{Wang2018Generalized} and \cite{Wang2017Block}, (G)SM is introduced to the NOMA system
 to improve both spectral efficiency and energy efficiency in the terrestrial return channel and uplink channel, respectively, in which the joint detection problem of user activity and (G)SM signals is explored by exploiting the block-sparse model.
NOMA-based SM is proposed in \cite{zhu2017NOMA} for downlink multi-user communications, which mitigates inter-user interference effectively via SIC and achieves high spectral efficiency at the cost of some performance degradation.
In \cite{Yang2019NOMA}, NOMA-aided precoded SM is proposed for downlink multi-user communications under overloaded transmissions, in which the demodulation of precoded SM is integrated into SIC to obtain
effective detection algorithms and analytical results in terms of spectral efficiency, implementation cost, multi-user interference, and
mutual information are also derived to evaluate the system performance.
In \cite{Wang2017Spectral} and \cite{Wang2017On}, NOMA-aided SM schemes with finite alphabet inputs are proposed for uplink and downlink communications, respectively, in which the lower bound on mutual information is derived to characterize the system's achievable rate.

It is noted that NOMA generally exploits inter-user channel strength disparities to
achieve an effective spectral utilization. Nevertheless, NOMA performance may degrade considerably
when different users have a similar channel gain. Noticing this issue, \cite{Kim2015Selective} proposes a switching mechanism between NOMA and SM, which maintains better spectral efficiency under minor inter-user channel strength disparities.
Against the switching mechanism, the combination scheme of NOMA and SM is developed
for uplink transmissions in \cite{Siregar2017Combination}, in which a hybrid
detection scheme is proposed to estimate the antenna index and then the constellation
symbols. In \cite{Chen2017Performance} and \cite{Chen2017NOMA-SM}, the combination scheme of NOMA and SM is proposed for vehicle-to-vehicle communications,
in which the index of one active antenna embeds one specific user's information
and the superposition coded symbol carried by the active antenna conveys the information for other users.

It is worth pointing out that in the works mentioned above, the implementation of superposition coding and SIC may incur a high computational cost as the number of users grows large. Moreover, interference cancellation can be imperfect in practice, which causes error propagation and further degrades the
system performance. To circumvent this issue, the study of \cite{Zhong2018Spatial} develops a primitive SM assisted
NOMA technique with antenna partitioning to support multi-user access, which avoids the utilization of SIC at the receiver and refrains from multi-user interference.
Furthermore, the SM-aided NOMA scheme is investigated under a three-node cooperative network in \cite{Li2019Spatial}, in which the information carried by
the index of one active antenna and the constellation symbol is separately delivered to two users to avoid the utilization of SIC as well as improve the system performance.
To support massive connectivity, SM aided code-division multiple-access is proposed as a NOMA-SM scheme for uplink communications in \cite{Liu2018Spatial}, whose performance and low-complexity signal detection problem are also investigated.

\subsection{Security provisioning in SM}
Due to the broadcast characteristics of the wireless interface, like other wireless transmission technologies, SM transmissions also face the security and information leakage challenges.
Specifically, the confidential messages delivered to legitimate users may be intercepted by malicious eavesdroppers.
As a result, ensuring secure transmission of SM becomes a critical topic in the application of wireless networks.
By complementing cryptographic techniques at the upper networking layers \cite{Zheng2018Learning}, PLS is an efficient alternative to improve the secrecy performance by exploiting the features of the wireless interface (e.g., randomness, variability, discrimination, reciprocity, etc.) \cite{Guan2012On}.
It is worth pointing out that from the physical layer perspective, both PLS and SM highly rely on the randomness and discrimination properties of the wireless interface.
Thus, an interesting problem arises when exploiting the channel properties to achieve confidential information exchange among legitimate nodes using SM techniques such that eavesdroppers fail to recover the information.
Therefore, advanced signal processing techniques may be applied to amplify the discrimination between the legitimate channel and wiretap channel, which benefits the legitimate node and impairs the eavesdropper at the same time.
Moreover, SM activates one antenna randomly for each channel use via a fast antenna-switching mechanism, which presents a fast time-varying environment to the received signal of the eavesdropper.
This information-driven antenna-switching mechanism makes it more difficult for the eavesdropper to intercept the modulation information carried on the randomly activated antenna, especially when the mapping rule is unknown to the eavesdropper.
Therefore, SM enjoys a great potential not only in the increased spectral efficiency and energy efficiency, but also in enhancing secure transmission.

In \cite{Sinanovic2012Secrecy} and \cite{Aghdam2015On}, some preliminary results on the secrecy capacity for both SSK and SM are presented, and the effects of the constellation size, received signal-to-noise ratio (SNR), and number of antennas at the transceiver on the secrecy behavior are examined.
Assuming that the CSI of the eavesdropper is available in the multiple-input-single-output (MISO) system, \cite{Guan2012On} studies the secrecy mutual information of SM under the finite input alphabet constraint and
develops a precoding based SM method to enhance the security performance.
Under the assumption of an available global CSI at the transmitter, the optimization problem of the precoding-aided SM system is investigated in \cite{Wu2016Transmitter} to improve the secure performance,
which maximizes the ratio of the received signal power at the legitimate user to that at the eavesdropper.
However, acquiring the eavesdropper's CSI may not be possible at the transmitter for passive eavesdroppers.
Without knowing the eavesdropper's CSI, artificial interference is introduced to the SM system for ensuring secure transmission in \cite{Wu2015Secret,Wang2015Secrecy,Chen2016Secure,Wu2018Performance}, where \cite{Chen2016Secure}
extends the precoding-aided SM technique to a multiuser downlink case against the multiple antenna eavesdropper.

In \cite{Huang2017Improving}, two random antenna selection methods are developed for precoding based SM to improve the secrecy rate, which severely degrades the reception quality of the passive eavesdroppers without degrading the received signal at destination.
In \cite{Shu2018Two}, two transmit antenna selection methods based on the maximization of
 signal-to-leakage-and-noise ratio and secrecy rate are developed for artificial interference-aided SM systems to enhance the transmission security.
A full-duplex legitimate receiver is introduced to the secure SM system in \cite{Liu2017Secure}, which not only receives the secret information from the source, but also emits jamming signals to cause time-varying interference at the passive eavesdropper.
Under the framework of QSM, an anti-eavesdropping scheme is proposed by transmitting mixed signals of the information symbol and the carefully designed artificial noise via randomly activated antennas \cite{Huang2017Anti-Eavesdropping}.
In \cite{Taha2018Secret}, by exploiting the randomness of the constellation mapping principle to encrypt the secret information, a novel scheme is presented for the secret key exchange and authentication in the SM system.

In \cite{Wang2016Spatial}, based on the channel reciprocity of time-division duplexing communications and knowledge of legitimate CSI,
the transmitter dynamically redefines the mapping rule of the spatial information to the active antenna indices for security purpose, and then chooses the active antenna according to the redefined mapping rule. Since a redefined mapping rule is unknown by the eavesdropper due to lack of the legitimate CSI, the security of the spatial information can be
guaranteed whereas the constellation information is not secure.
Attentive to the security issue of the constellation information,
\cite{Jiang2018Secrecy-Enhancing} dynamically rotates the active antenna indices and the constellation symbols at the transmitter according to
the instantaneous legitimate CSI, which is totally unknown to the eavesdropper.
As a result, both the spatial information and constellation information are secure from
the eavesdropper.
Moreover, by leveraging the random state of the legitimate channel, the mapping-varied SM scheme is proposed in \cite{Yang2018Mapping-Varied}, which dynamically rearranges the mapping patterns of the antenna index bits and constellation bits according to the instantaneous legitimate channel quality. Without knowing the legitimate channel state, the eavesdroppers are confused with variable information mapping patterns and the PLS can be further enhanced.

The design of the SM technique for secure cooperative communication systems is also investigated in \cite{Gao2018Secure} and \cite{Bouida2017Precoded}.
In \cite{Gao2018Secure}, a securely distributed SM scheme is deployed in multiple non-dedicated relays to defend against eavesdropping attacks and coordinate the transmission of source-destination information via the indices of the selected relays.
In particular, the selected relays are equipped with specifically designed precoder to facilitate the decoding of spatial information at the legitimate receiver while confusing the eavesdropper.
In \cite{Bouida2017Precoded}, a precoded SM-based cooperative network using relay selection and jamming is investigated with secrecy considerations, where Alice applies precoded SM and Bob emits a jamming signal simultaneously for security protection in the first phase while multiple relays coordinate the information transmission and jam the eavesdroppers in the second phase. Moreover, the power allocation optimization problem among the cooperative relays and the performance in terms of secrecy capacity and outage probability are studied.

\section{Applications of SM to Emerging Communication Systems} \label{Applications SM}
\subsection{SM in mmWave Communications}
Driven by the high cost and exhaustion of radio resources in low frequency bands (below 6 GHz), the exploitation of mmWave frequency band (around 60 GHz) alleviates the data traffic burden and
provides super high data services in a cost-efficient manner \cite{Rangan2014Millimeter-Wave,wu2019beef}.
However, working in high-frequency bands, mmWave communications suffer from high propagation loss. To overcome this drawback, beamforming technology can be applied with a large number of antennas to compensate for high propagation attenuation.
Fortunately, attributed to the significantly reduced wavelength in the mmWave frequency band, mmWave MIMO systems can be equipped with a large number of antennas at the transceiver in a highly compact manner to achieve high beamforming gains \cite{Hur2013Millimeter}.
For the conventional mmWave MIMO system, the high compaction of antennas imposes the prohibitively high cost at the transceiver due to the use of a large number of RF circuits, which hinders the practical implementation.
To explore the practical implementation of RF-chain-limited mmWave systems,
the (G)SM technique is a promising low-cost and high-efficiency design option by activating a subset of transmit antennas connected to a small number of RF chains.
As a result, (G)SM becomes an attractive technology in mmWave communications for reaping  benefits of both beamforming gain and low cost with reduced RF chains.

\begin{figure}[!t]
	\centering
	\includegraphics[width=3.5in]{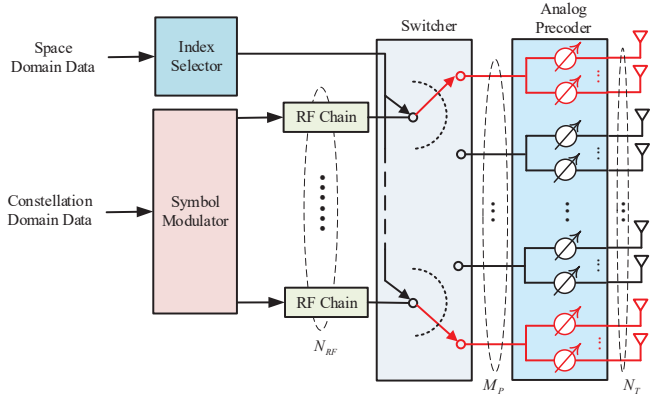}
	\caption{Transmitter diagram for (G)SM aided mmWave communication systems.}
	\label{gsmmmwave}
\end{figure}
Fig.~\ref{gsmmmwave} shows the transmitter diagram of the (G)SM aided mmWave communication systems. A total number of $N_T$ transmit antennas are partitioned into $M_P$ analog precoders, each of which comprises $N_K$ antenna elements for analog precoding, i.e., $N_T=M_P\cdot N_K$. For the implementation of (G)SM aided mmWave scheme, the number of RF chains
$N_{RF}$ is less than the number of analog precoders $M_P$.
Unlike the conventional mmWave MIMO scheme that conveys information in the constellation domain,
the incoming data of (G)SM aided mmWave scheme are separated into two parts in the constellation domain and space domain, respectively.
Specifically, the space domain data are fed to the index selector to determine the indices of $N_{RF}$ out of $M_P$ analog precoders for activation.
Since the remaining $M_P -N_{RF} $ analog precoders are set to be idle,
 the outputs of $N_{RF}$ RF chains are connected with $N_{RF}$ activated analog precoders via a fast switcher and the total combination number of $N_{RF}$ analog precoders is ${{M_P}\choose{N_{RF}}}$. On the other hand, the constellation data are fed to the symbol modulator to generate $N_{RF}$ constellation symbols, and then independently processed with $N_{RF}$ RF chains. For each analog precoder, $N_K$ antenna elements are deployed as phase shifters for the analog procession. A diagonal matrix of size $N_T$ is applied as an analog precoder, whose $n$-th diagonal entry is given by
\begin{align}\label{analog precoder}
\left[{\bf A}\right]_{n,n}=\frac{1}{\sqrt{N_K}}e^{j \phi_n}
\end{align}
 where $\phi_n \in [-\pi, \pi]$ is the rotation angle of the $n$-th transmit antenna.
 Similar to (\ref{GSM transmission block}), we denote the (G)SM block as ${\bf x} \triangleq \left[x_1~ x_2 \ldots x_{M_P}\right]^{T} $, and the received signal model can be expressed as
 \begin{align}\label{rec_mmWave}
 {\bf{y}} &=\sqrt{\frac{\rho}{K}} {\bf{HA }} \left({{\bf x} \otimes {\bf 1}_{N_K}}\right)  + {\bf{w }}
 \end{align}
 where ${\bf 1}_{N_K}$ denotes the all-one vector of size $N_K$, $\otimes$ denotes
 the Kronecker product, ${\bf{A}}$ is the mmWave MIMO channel matrix of size $N_R \times N_T$, and ${\bf{w }}$ is an AWGN vector at the receiver.
 Note that when $N_{RF}=1$, the system in (\ref{analog precoder}) specializes to a
 SM aided mmWave system with only a single RF chain, which enjoys an extremely low cost of implementation and detection in mmWave communications.
(G)SM aided mmWave scheme not only saves the transmitter cost by reducing the number of RF chains, but also achieves the precoding gains via the analog precoder, which meets the requirement of mmWave communications.

In \cite{Ishikawa2017Generalized-Spatial-Modulation-Based}, the implementation of analog beamformers in (G)SM-based mmWave systems is investigated, where the array of analog beamformers is optimally designed to attain the maximum rank of the equivalent channel matrix.
Moreover, it is shown that with a reduced RF-chain structure, the (G)SM-based
mmWave scheme of \cite{Ishikawa2017Generalized-Spatial-Modulation-Based} approaches the same constrained capacity as the full-RF spatial-multiplexing
counterpart.
The application of SSK and (G)SM in indoor mmWave communications at 60 GHz is studied in \cite{Liu2015Space,Liu2016Line-of-Sight,Liu2018Performance}, where the channel is characterized as line-of-sight (LOS) components due to the high reflection attenuation. 
Based on the maximum criterion of the minimum distance of received symbols, SSK \cite{Liu2015Space} and (G)SM \cite{Liu2016Line-of-Sight,Liu2018Performance} in LOS channels are designed to achieve optimal performance by orthogonalizing the channel matrix with proper antenna placement.
By approximating the 3-D statistical channel as a log–normal fading channel, \cite{Younis2017Quadrature} analyzes the capacity of QSM for outdoor mmWave communications.
Assuming the CSI at the transmitter, \cite{He2017Spectral-Efficient} proposes an analog precoding aided (G)SM scheme for
mmWave communications to improve the spectral efficiency. Specifically, after deriving a tight lower bound on the achievable spectral efficiency, the analog precoding aided (G)SM scheme
in \cite{He2017Spectral-Efficient} is designed in an iterative way to meet the maximal spectral efficiency. Furthermore, hybrid precoding aided (G)SM schemes using hybrid
digital and analog precoding are proposed for mmWave communications, which achieve an optimal
spectral efficiency based on some approximations \cite{He2017On,He2018Spatial}.

To save the transmitter cost with a reduced number of antennas,
(G)SM with a variable number of activated antennas is designed in \cite{Liu2017Variable} for indoor LoS mmWave communications to achieve higher transmission efficiency.
Moreover, the capacity and energy efficiency are analyzed,
and the optimization problems in terms of power allocation and antenna separation are investigated.
To save energy consumption and allow robustness against path-loss attenuations, an orbital angular momentum aided SM scheme is proposed for mmWave communication systems in \cite{Ge2017Millimeter}, where the
capacity, error performance, and energy efficiency are also discussed.
In \cite{Yang2017Adaptive}, a SM aided receive antenna selection technology is developed for saving the RF cost at the transceiver in mmWave communications, where the combinatorial optimization problems of
capacity and error performance are studied.
To attain beamforming gain and reduce the cost of RF chain, an analog precoding-aided virtual SM \cite{Lee2016Transmitter} and its variant using multi-mode hybrid precorder \cite{Lee2017Adaptive} are proposed, which enhance the received SNR and spatial degrees of freedom with the low cost of single RF chain.
Receive SM is also studied in indoor and outdoor mmWave communications \cite{Perović2017Receive,Raafat2017Receive,Zhu2018Low}, where the performance is optimized with minimum symbol error probability in LOS channels \cite{Perović2017Receive}.

\subsection{SM in Optical Wireless Communications (OWC)}\label{VLC}
In addition to mmWave communications discussed in the last subsection, OWC is another appealing complement to alleviate the shortage problem of wireless radio resources and achieve high
data rate transmission \cite{Sevincer2013LIGHTNETs}, which can be categorized into indoor and outdoor implementations depending on the communication range.
For example, by deploying low-cost LEDs for the indoor environment, OWC can provide simultaneous energy-efficient lighting and high throughput transmission.
As an efficient and inexpensive dual-use technology of illumination and communications, OWC enjoys a lot of advantages in both indoor and outdoor communications, such as cost efficiency, license-free implementation, large bandwidth, and high reliability.

The primitive optical SM is developed in \cite{Mesleh2010Indoor} for the indoor environment, which can be extended to various scenarios.
In \cite{Ozbilgin2015Optical}, an optical SM is developed for outdoor optical communications to achieve a higher spectral efficiency and energy efficiency,
where the analytical and simulation results of both uncoded and coded systems are presented under the turbulence induced fading.
Based on coherent detection, a theoretical framework for error performance analysis is presented for both uncoded and coded optical SM systems over atmospheric turbulence characterized by HK distribution in \cite{Peppas2015Free-Space}, which shows enhancement in performance, spectral efficiency and deployment cost.
Under the consideration of atmospheric turbulence and pointing error,
the error performance and mutual information of the optical SSK are studied over three different optical channel models \cite{Jaiswal2018An}.
Considering the inter-symbol interference incurred by delay spread, \cite{Olanrewaju2018Performance} studies the error behavior of the optical SM over indoor multipath optical channels, whose upper bound on error performance is also derived.
 With CSI available at the transmitter, \cite{Wang2017Constellation} proposes an optical SM based on the constellation
optimization, whose minimum Euclidean distance is maximized at the receiver.
To deal with the adverse effect on the detection of optical SM incurred by the high channel correlation, an optical SM based on the power imbalance is proposed for OWC in \cite{Ishikawa2015Maximizing}, which optimizes the constrained capacity using power allocation.
In \cite{Fath2011Spatial}, the error performance of the SM aided OWC system is investigated over the indoor environment characterized by LOS components, which outperforms the repetition coding aided OWC system.
In \cite{Wang2016On}, the mutual information and its corresponding lower
bound of the SM aided OWC system are theoretically calculated
 under the MISO and finite input alphabet setting. Based on these theoretical results, the optimal precorder is designed to maximize the minimum Euclidean distance.
 An enhanced optical SM scheme is proposed for indoor OWC in \cite{Mesleh2011Performance}, which aligns multiple transmit-receive units to achieve higher SNR gains.
 Under the consideration of imperfect synchronization at the transmitter, the theoretical error performance of several variants of optical SM is derived for OWC in \cite{Olanrewaju2017Effect},
 which provides some insightful information about the effect of synchronization
 error.

For indoor environments, optical wireless communications can be deployed as VLC for simultaneous luminance and communication.
In particular, as one indoor implementation of OWC, VLC is an attractive alternative to apply in some RF-free working environments, e.g., hospital, airplane, and gas station \cite{Karunatilaka2015LED,Popoola2016Impact}.
Due to the wide application of LED array comprising a large number of LEDs for sufficient illumination, MIMO technology can be readily integrated into VLC to achieve high transmission rates, where various representative MIMO aided VLC schemes have been investigated and compared in \cite{Fath2013Performance}.
Thanks to its capability of supporting fast switching in LEDs, SM aided VLC can be deployed, which activates a single LED only in the LED array for each channel use to convey index information additionally through the location of the active LED.
Moreover, by adjusting the number of active LEDs and the intensities, both (G)SSK and (G)SM provide the flexibility to control the luminance of the LED array and the communication throughput.
To attain higher power efficiency, an enhanced optical (G)SM scheme based on the collaborative constellation pattern is developed for indoor VLC in \cite{Kumar2017Power,Kumar2018Dual-Mode,Kumar2018Improved}.
Using the ML-based photodetectors, the theoretical and experimental results of (G)SSK are presented for the short-range indoor VLC in \cite{Popoola2013Error} and \cite{Popoola2014Demonstration}. The error performance of (G)SM aided VLC systems is presented with an analytical upper bound on BER in \cite{Alaka2015Generalized}. Exploiting available CSI at the transmitter for interchannel interference reduction, an optical SM based on channel adaption is designed to apply in massive MIMO-VLC in \cite{Xu2016Channel-Adapted}.
Moreover, the combination of optical SM and layered
space-time coding is designed for VLC by mapping the spatial information into different space-time coding matrices. To achieve the control flexibility on luminance and the data throughput, a (G)SM aided
hybrid dimming strategy using clipped optical OFDM is designed for VLC in \cite{Wang2018Spectral-Efficient}, whose achievable rate is theoretically analyzed over channels with dense scattering and high correlation, respectively.
To achieve a flexible design with an arbitrary number of transmit antennas for SM,
a three-domain bit mapping solution based on channel adaptation is suggested in \cite{Wang2018Adaptive}, whose lower bound on achievable rate is also obtained in closed-form.
Noticing the real and positive constraint on light intensity levels in intensity modulation/direct detection system, the authors of \cite{Yesilkaya2017Optical} and \cite{Başar2016Generalized} successfully apply the (G)SM principle to the optical MIMO-OFDM system, which exploits the advantage of SM based LED to avoid Hermitian symmetry and direct current (DC) bias.
Specifically, each complex OFDM signal are split into four components consisting of the real-imaginary and positive-negative parts and then encoded by four LEDs for transmission, which dispenses with DC biasing/asymmetrical clipping \cite{Yesilkaya2017Optical}.

\subsection{Hardware Implementation of SM}\label{hardware}
It is worth pointing out that most research works on SM make the impractical assumption of ideal hardware at the transceiver.
Therefore, practical implementation of the SM family is an important and challenging research direction, since the channel estimation, transceiver movement, and hardware impairments due to phase noise, I/Q imbalance,
and nonlinearity of the amplifier can have negative effects on the system performance.
The first attempt on the hardware implementation and experimental
assessment of the SM principle is reported in \cite{Serafimovski2013Practical}, where the
testbed platform with two National Instruments (NI)
PXIe devices is designed to verify the experimental performance of SM under the indoor propagation environment characterized by LoS wireless channels.
In particular, the implementation and applicability of SM has been first approved, which reveals the impact of various hardware impairments at the transceiver, e.g., imperfect channel estimation, spatial correlation, and I/Q imbalance.


On the other hand, it is preferable to use low-resolution analog-to-digital converters (ADCs) with few-bits quantization at the user terminal to achieve a much lower power consumption and reduce hardware cost significantly. In \cite{Wang2015Multiuser}, the multiuser detection problem of the SM system is investigated for the large-scale MIMO implementation employing low-resolution ADCs. It is shown that the SM system developed in \cite{Wang2015Multiuser} is robust to quantization error and channel correlation, and the proposed low-complexity detector achieves better performance than its linear counterparts.
Noted that the antenna-switcher incurs time delays and is not cost-free, \cite{Soujeri2016The} studies the impact and limitation of the antenna-switching time for SM, in which the spectral efficiency and achievable rate are analyzed under the practical hardware constraint of antenna-switching.
In \cite{Afana2017Performance} and \cite{Afana2018On}, the performance of SM based cognitive radio communications is investigated under the consideration of channel estimation error and hardware imperfections, where some theoretical results are derived for Rayleigh fading channels.

The hardware requirements and limitations of various SM members are discussed in \cite{Mesleh2017Transmitter}, which shows that hardware impairments have a significant influence on practical performance.
After developing an analytical framework under the transmit I/Q
imbalance for SM,
the authors of \cite{Helmy2017On} show the robustness of SM to the I/Q imbalance.
In \cite{Lee2016Implementation}, a practical hardware implementation of the SM detector is realized by using 87.4-K logic gates in the 0.18-$\mu$m CMOS technology, which adopts the signal-vector-based list detection and dual-data-path architecture to achieve low hardware complexity and near-optimal performance.
Furthermore, hardware realization of the ML detection is implemented with 70.5-K gates under the TSMC 90-nm CMOS technology in \cite{Liu2019A}, where the Givens rotation is applied to achieve stable division operation and low-complexity implementation.



\subsection{SM Based Simultaneous Wireless Information and Power Transfer (SWIPT)}
	In all works mentioned above, the inactive antennas of an SM node are either kept idle or turned on for signal reception when operating in full-duplex mode. With the development of SWIPT technology, researchers begin to realize a new energy harvesting function of the inactive antennas in SM systems. SM based SWIPT has been therefore receiving considerable attention recently. Since the research topic is relatively new, there are only a few works in the literature. In summary, they can be still classified into three categories.
	
	In the first category, the SM signal is exploited for SWIPT by power splitting \cite{Guo2016Spatial}, \cite{Koc2018Two-Way}, \cite{Zhang2015Energy}, \cite{Cheng2017On}. Specifically, in \cite{Guo2016Spatial} a power splitting-based SWIPT is considered for an SM system with a multiple-antenna receiver. The authors assume different power splitting factors for different receive antennas and develop an iterative algorithm to find the optimal values in maximizing the achievable rate under certain energy harvesting constraints. \cite{Koc2018Two-Way} develops a dual-hop relaying network using SM at both sources and physical-layer network coding, where the relay obtains required energy from the RF SM signals via the power splitting protocol. Both full-duplex and half-duplex modes are considered for the network, and the full-duplex mode is shown to outperform its half-duplex counterpart as the quality of self-interference cancellation and/or the spectral efficiency increases. While \cite{Guo2016Spatial} and \cite{Koc2018Two-Way} consider transmit SM, \cite{Zhang2015Energy} and \cite{Cheng2017On} shed light on the application of receive SM to SWIPT. In \cite{Zhang2015Energy}, an energy pattern aided SWIPT system is proposed, where the information is conveyed not only by the indices of the activated receive antennas but also by the specific intensity of the delivered power. Results on BER demonstrate the beneficial immunity of the proposed system to power conversion. Unlike \cite{Zhang2015Energy} that focuses on joint design, \cite{Cheng2017On} proposes to transmit a separated energy and receive SM signals, such that the predefined energy signal can be removed at the information receiver to facilitate signal detection.
	
	In the second category, the SM node participates in both information decoding and energy harvesting \cite{Zhang2017Spatial}, \cite{Qu2019Full}. This idea was first coined in \cite{Zhang2017Spatial}, where a GSM node communicates to an information receiver through the active antennas while harvesting energy transmitted from an energy transmitter and recycling part of its self-energy through the inactive antennas. Enhancement in terms of achievable rate over the existing time switching-based and antenna selection-based wireless powered communication protocols is verified therein. Later, the idea was extended to a dual-hop relaying network in \cite{Qu2019Full}, where a GSM-based full-duplex relay without an external power supply is assumed, and three different realizations were proposed. While both the first and second methods are based on time switching, the third method is based on power splitting. The first method performs energy harvesting first, suffering from self-interference in the following information decoding step, while by reversing the order the second method turns the adverse self-interference into an advantage to energy harvesting as \cite{Zhang2017Spatial}. The third method carries out information decoding and energy harvesting simultaneously all the time, improving the transmission efficiency.
	
	In the third category, the SWIPT capability is developed for distributed SM \cite{Narayanan2018Wireless}. Assuming multiple geometrically separated single-antenna relays to form distributed SM, the authors of \cite{Narayanan2018Wireless} propose two different protocols to enable SWIPT. The first protocol allows each relay node to harvest energy transmitted from the source, which is used by the activated relay to forward the source's and its own data simultaneously. The second protocol exploits the inactive relays to recycle part of the transmitted energy in the network in addition to harvesting the transmitted energy from the source. Both protocols retain the original properties of DSM, but the second protocol enjoys a lower hardware complexity and better error performance than the first protocol in certain scenarios.

%
%
%
%
%
%

\section{Applications of SM in New Domains}\label{Extension SM}
\begin{table*}[!t]
	\begin{center}\caption{Classification of Representative Forms of SM.}\label{Table of Classification}
\begin{tabular}{m{40pt}<{\centering}|p{40pt}<{\centering}|m{50pt}<{\centering}|m{120pt}|m{160pt}}
	 \hline
	Type &     \multicolumn{ 2}{|c|}{Entity} & Representative Schemes & Main Achievements \\
	\hline
	\hline
	{} &    {} & Antenna & SM\cite{Mesleh2008Spatial}, SSK\cite{Jeganathan2009Space}, (G)SM\cite{Younis2010Generalised}, (G)SSK\cite{Jeganathan2008Generalized}, QSM\cite{Mesleh2015Quadrature} & Higher energy efficiency \& lower deployment cost \\
	\cline{3-5}
	{} &   Space Domain     &  LED &        OSM\cite{Mesleh2011Optical} & Flexibility to control illumination \& communications \\
	\cline{3-5}
	{} &  {} & {RF Mirror} &  RA-SSK\cite{Bouida2016Reconfigurable}, CM\cite{Basar2017Space-Time}& Better system performance
	\& lower antenna cost \\
	\cline{2-5}
	{} & Frequency Domain & Subcarriers &
	SIM-OFDM \cite{Abu-alhiga2009Subcarrier},
	 OFDM-IM\cite{Basar2013Orthogonal}, OFDM-I/Q-IM\cite{Zheng2015Low-Complexity}, GFDM-IM\cite{Ozturk2016Generalized} & Higher spectral efficiency\&  higher energy efficiency  \\
	\cline{3-5}
	{Single-Dimension Entity} & {} & Subcarriers & CI-OFDM-IM\cite{Başar2015OFDM}, MIMO-OFDM-IM\cite{Başar2015Multiple-Input,Basar2016On,Zheng2017Multiple-Input,Zheng2017Low-complexity}& Higher reliability\&  higher energy efficiency \\
	\cline{2-5}
	{} &      Time Domain  & Time slot &      SC-IM\cite{Nakao2017Single-Carrier} & Higher transmission efficiency for broadband systems \\
	\cline{2-5}
	{} &     Code Domain   & Spreading Code & CIM-SS\cite{Kaddoum2015Code}, GCIM-SS\cite{Kaddoum2016Generalized}, IM-OFDM-SS\cite{Li2018Index} & Higher spectral efficiency \& lower energy consumption  \\
	\cline{3-5}
	{} & {} & Modulation Type& ESM\cite{Cheng2015Enhanced}, DM-OFDM\cite{Mao2017Dual-Mode}, MM-OFDM-IM\cite{Wen2017Multiple-Mode} & Higher spectral efficiency \\
	\cline{2-5}
	{} &     Angle Domain   & Polarization state& PolarSK\cite{Zhang2017Polarization}, 3-D PMod\cite{Henarejos20183-D} & Higher spectral efficiency \& lower hardware cost \\
	\cline{3-5}
	{} & {} & AoA & BACM\cite{Hoseyni2017Beam} & Higher spectral efficiency \\
	\hline
	{} & \multicolumn{ 2}{|c|}{Space-Time} &   STSK\cite{Sugiura2010Coherent}         & Flexible tradeoff between diversity \& multiplexing gain  \\
	\cline{2-5}
	{Multi-Dimension Entity} & \multicolumn{ 2}{|c|}{Space-Frequency} &  GSFIM\cite{Datta2016Generalized} & Better system performance \\
	\cline{2-5}
	{} & \multicolumn{ 2}{|c|} {Space-Time-Frequency} &   STFSK\cite{Ngo2011Space-Time-Frequency}, GSTFIM\cite{Kadir2018Generalized}       & Higher spectral efficiency \\
	\hline
		\end{tabular}
\end{center}
\end{table*}

As a novel digital modulation technology for achieving both spectral efficiency and energy efficiency, the basic concept of SM is to convey additional information through the ON/OFF states of transmit antennas. Consequently, the concept of SM can be generalized and used in various applications, in which the transmission media for embedding additional information
can be either physical (e.g., antenna) or virtual (e.g., space-time matrix).
Instead of using the full-activation state, the distinguishing feature of SM is to embed the {\it implicit} information via the activation states of transmission media in addition to
{\it explicit} information (e.g, constellation symbols), which is carried by the partially activated transmission media \cite{Basar2016Index,Cheng2018Index,Wen2017Index,Basar2017Index,Yang2018Multidomain}. The information loss due to the deactivation of some transmission media can be compensated for by the implicit information based on the ON/OFF states,
saving the cost of transmission media (the number of actual transmission media is significantly reduced) and striking an attractive compromise between energy efficiency and spectral efficiency.

Relying on the activation states of transmit antennas, SM provides a new perspective on the digital modulation, which is different from the conventional digital modulation using the amplitude, frequency, and phase on the sinusoidal carrier.
Inspired by the concept of SM, researchers have proposed various forms and applications for SM, which alter the ON/OFF states of different transmission entities
to embed additional information.
In Table \ref{Table of Classification}, we summarize extensive representative forms of SM, which are classified according to their transmission entities that perform the ON/OFF keying mechanism. As Table \ref{Table of Classification}, the concept of SM can be performed in either a single-dimension entity or a multi-dimension entity.
So far, various single-dimension entities have been explored to
increase the spectral efficiency and energy efficiency of communication systems,
which mainly includes spatial entities (e.g., antenna, RF mirror, and LED), frequency entities (e.g., subcarrier), time entities (e.g., time slot), code entities (e.g., spreading code and modulation type), angle entities (e.g., AoA and polarization state), etc.
To further enhance the system performance and enjoy a more flexible system setting, several multi-dimension entities are also developed, which include more than one dimension for performing the ON/OFF keying mechanism.
\subsection{Single-Dimension Entity}
After the pioneering works in \cite{Mesleh2008Spatial} and \cite{Jeganathan2009Space}, the emergence of
SM and SSK schemes has sparked renewed and widespread interest in digital modulation, both of which require a single RF chain only and embed spatial information through the active
antenna index.
Afterwards, different variants of the SM technology have been developed to accommodate more than one active antennas and further enhance the spectral efficiency, which include (G)SM \cite{Younis2010Generalised}, (G)SSK\cite{Jeganathan2008Generalized} and QSM \cite{Mesleh2015Quadrature}.
Besides the spatial entity of antenna, the implementation of SM can also be extended to communication systems deploying RF mirrors and LEDs.
As presented in Section~\ref{VLC}, the SM aided VLC, which conveys spatial information through the location of the active LED, is an efficient and inexpensive dual-use technology of illumination and communications \cite{Mesleh2011Optical}.
Another remarkable implementation of SM in the space domain is
the reconfigurable antenna (RA) based system, which controls the ON/OFF states of the
RF mirrors surrounding a transmit antenna and further alters the radiation
pattern to embed additional information \cite{Renzo2017Spatial,Phan-Huy,viet2019spatial,Phan-Huy2017First}.
With the adjustable radiation characteristics of the transmit antenna, RA-SSK is proposed in \cite{Bouida2016Reconfigurable} to improve the system performance
and reduce the implementation cost in transmit antennas.
Moreover, as the changing of radiation pattern results in the variations of the channel characteristics,
it can be regarded as the modulation based on the channel state, which is also referred to as channel modulation (CM) schemes \cite{Basar2017Space-Time}.

Although the SM technology originated in the space domain, SM is not exclusive to the space domain and can be easily transplanted into the frequency/time/code/angle domains.
Inspiring by the SM concept, the authors of \cite{Abu-alhiga2009Subcarrier} introduce
	 the subcarrier-index modulation (SIM)-OFDM system,
which is the first attempt to exploit the subcarrier indices for encoding information and alleviating the high peak-to-average power ratio issue in OFDM systems.
By replacing the transmission of antennas with subcarriers for performing the ON/OFF keying mechanism,
orthogonal frequency division multiplexing with index modulation
(OFDM-IM) is proposed in \cite{Basar2013Orthogonal} as a successful and inspiring representative for implementing the SM concept into multi-carrier systems.
Inspired by the design of the QSM scheme, OFDM with in-phase/quadrature index modulation (OFDM-I/Q-IM)
is proposed in \cite{Zheng2015Low-Complexity} to perform the ON/OFF keying mechanism
independently in the in-phase and quadrature dimensions, respectively, which
further enhances the spectral efficiency and the error performance.
In \cite{Fan2015Generalization}, the generalization of OFDM-IM is proposed to further enhance the spectral efficiency and flexibility with a variable number of active subcarriers, which
relaxes the constraint of the fixed number of active subcarriers in \cite{Basar2013Orthogonal}.
The concept of SM is also applied to generalized frequency division multiplexing (GFDM) in \cite{Ozturk2016Generalized}, which yields the GFDM-IM scheme and achieves performance gain over conventional GFDM.
To achieve higher diversity gain, a novel coordinate
interleaved OFDM-IM (CI-OFDM-IM) scheme is proposed in \cite{Başar2015OFDM} by combining the SM concept and coordinate interleaved STBC in OFDM systems.
The combination scheme of OFDM-IM and MIMO, which referred to as MIMO-OFDM-IM,
 is investigated in \cite{Başar2015Multiple-Input} to achieve higher data rates and better error performance than the traditional MIMO-OFDM scheme, where various low-complexity and near-optimal algorithms are developed for the demodulation of
 MIMO-OFDM-IM \cite{Başar2015Multiple-Input,Basar2016On,Zheng2017Multiple-Input,Zheng2017Low-complexity}.

The successful implementation of OFDM-IM
has also propelled the spread of the SM concept to the time, code, and angle domains.
In \cite{Nakao2017Single-Carrier}, a novel single-carrier (SC) based
IM (SC-IM) scheme, which activates a subset of symbols in each time-domain subframe to convey additional information, is proposed to enhance the transmission efficiency
for broadband systems.
A code index modulation-spread
spectrum (CIM-SS) is proposed in \cite{Kaddoum2015Code} to achieve better performance with reduced energy consumption,
which carries additional information by choosing one of the two spreading Walsh codes.
Moreover, a generalization scheme of CIM-SS (called GCIM-SS) is then proposed in \cite{Kaddoum2016Generalized} to further improve the spectral efficiency, which increases the set size of the spreading Walsh codes to bear more index information.
By integrating the SM concept into OFDM spread spectrum (OFDM-SS) systems,
the authors of \cite{Li2018Index} propose an index modulated
OFDM-SS (IM-OFDM-SS) scheme to harvest the diversity gain and higher spectral efficiency, which embeds the index information through the selection of spreading codes and can be applied to multi-user systems.
By enabling the flexibility of activating one or two antennas, an enhanced SM scheme is developed in \cite{Cheng2015Enhanced} to overcome the compromised spectral efficiency of conventional SM,
which jointly select the active antenna(s) and the constellations for embedding more index information.
To mitigate the loss in spectral efficiency due to the inactive subcarriers,
a dual-mode OFDM (DM-OFDM) is proposed in \cite{Mao2017Dual-Mode} to enhance the
data transmission efficiency, in which the originally active subcarriers of OFDM-IM carry the modulated symbols drawn from the primary constellation while
those inactive subcarriers of OFDM-IM are reactivated to carry new modulated symbols drawn from a secondary constellation.
Furthermore, the generalization scheme of DM-OFDM, which is referred to as
multiple-mode OFDM-IM (MM-OFDM-IM) \cite{Wen2017Multiple-Mode} and includes more constellation modes, is developed
by exploiting the full permutation pattern to carry more index information.
By exploiting polarization states for conveying index information through the angle domain, a generalized polarization shift keying (PolarSK) modulation scheme, is proposed for dual-polarization communication systems in \cite{Zhang2017Polarization} to achieve high spectral efficiency while retaining a low hardware cost.
Moreover, a 3-D polarized modulation (PMod) is developed for dual-polarization communication systems in \cite{Henarejos20183-D}, which selects both polarization state and the radiation phase according to the information bits to achieve better error performance.
In \cite{Hoseyni2017Beam}, a beam angle channel modulation (BACM) scheme is introduced to convey index information in the angle domain, which embeds the information into the AoA of a transmission beam.


\subsection{Multi-Dimension Entity}
As discussed in the last subsection, various applications based on the SM concept have been implemented in the space/frequency/time/code/angle domains, which exploit the ON/OFF states or the permutations of transmission entities in one domain to carry index information.
To further enhance the system performance and enjoy a more flexible system setting,
it is possible and beneficial to integrate resources across multiple dimensions to form a multi-dimension entity for the implementation of SM concept.

To unify the SM concept in both space and time domains,
both coherent and differential STSK modulation schemes are developed via the activation mechanism on selecting one out of multiple space-time dispersion matrices, which enables a flexible design to compromise between diversity and multiplexing gains \cite{Sugiura2010Coherent}.
By extending the SM concept to include both space and frequency dimensions, the authors of \cite{Datta2016Generalized} develop a generalized space
and frequency IM (GSFIM) scheme, which shows
the potential to achieve better performance than the traditional MIMO-OFDM scheme.
Exploiting the ON/OFF keying mechanism across the space, time and frequency domains, space-time-frequency shift keying (STFSK) scheme is also proposed in \cite{Ngo2011Space-Time-Frequency} to reap the potential gain from the three domains, which shows an improvement in the system performance and
 alleviation of inter-symbol interference over frequency-selective fading channels.
 Based on the dispersion matrix activation over space, time
 and frequency domains, a generalized space-time-frequency index modulation
 (GSTFIM) scheme is proposed in \cite{Kadir2018Generalized} to provide extra spectral efficiency.

\section{Conclusions and Future Directions}\label{Conclusions}
SM is a promising digital modulation technology to fulfill the requirements of emerging wireless systems due to its potential of achieving high energy efficiency, low deployment cost, free
of interchannel interference, relaxed inter-antenna synchronization requirements,
and compatibility with massive MIMO systems.
As shown in this survey, relying on the activation states of transmit antennas to convey additional information, SM can achieve an attractive compromise between spectral efficiency and energy efficiency with simple design philosophy, which has been verified by extensive studies.
We have first discussed the basic principles, variants, and enhancements of SM, and then shown the broad prospects of the SM concept in various implementations, including
integration with other promising techniques, applications to emerging communication systems, and extensions to new domains.
We hope that this survey and the research results in this special issue will be helpful to the readers to gain a better understanding and clearer picture on the
advantages and opportunities of the SM family as well as its wide-range applications.

Although considerable significant work has been studied on the research filed of SM, there are still many interesting as well as challenging
research problems of SM to be tackled in order to further broaden its applications and fulfill the requirements of future wireless communications.
Therefore, we close this survey with some promising directions worthy of investigation in future research, which are summarized as follows:
\begin{itemize}
	\item The scalability and integration of SM techniques to multi-user networks. In particular, scheduling and resource allocation of SM based-networks is one of the indispensable tasks to maximize the system throughput and balance user fairness. Moreover, the antenna and power allocation problem of SM should be further investigated to achieve an effective multi-user interference management.
	\item The high-mobility management problem of the SM technology in dynamic and mobile networks such as vehicle-to-everything (V2X) communications, unmanned aerial vehicles (UAV) communications, and underwater acoustic (UWA) communications. For example, since the occupation of pilots consumes a non-negligible fraction of space/time/frequency/code resources, the number and pattern of pilots have a significant impact on the high-mobility SM system, which should be carefully designed. Moreover, similar to the concept of SM, the distinguishable pilot patterns may be exploited to carry additional information bits and compensate for the overhead cost of the pilot occupation in the high-mobility SM system.
	\item Although SM is a promising candidate for massive MIMO, the channel acquisition/estimation can be a challenging task due to the massive number of antennas. In fact, the required channel training overhead is proportional to the number of antennas, which may be prohibitive or even unaffordable in practice. Furthermore, the channel feedback problem becomes even more prominent and challenging to the link-adaptive SM system under the massive MIMO setup.
	\item Note that the error performance of the spatial information bits and the constellation information bits of SM can be significantly different under various wireless environments. This is attributed to the fact that the error performance of spatial information bits highly depends on the differentiability of the channel signatures.
	Therefore, how to balance the error performance difference and schedule the information bits is an essential problem of SM.
	\item Relying on the ON/OFF state for embedding information, SM can be an energy efficient and cost-effective technique for the IoT to connect a massive number of machine-type communication (MTC) devices. Since MTC devices are usually low data rate requirements, periodic data traffic arrivals, limited signal processing capability, and strict energy constraints, the specific design and integration of SM to the IoT is also a promising research direction.
	\item The fundamental trade-off problem of SM between energy/spectral efficiency and the other communication requirements such as high reliability, high capacity, flexible design, and low latency.
	 \item The application/generalization of the SM concept to new emerging transmission entities such as orbital angular momentum can also bring new challenging problems.
\end{itemize}

\ifCLASSOPTIONcaptionsoff
  \newpage
\fi



\end{document}